\documentclass[twocolumn,showpacs,superscriptaddress,preprintnumbers,amsmath,amssymb]{revtex4}
\usepackage{graphicx}% Include figure files
\usepackage{dcolumn}% Align table columns on decimal point
\usepackage{bm}% bold math
\usepackage[perpage]{footmisc}

\begin{document}

%\setlength{\topmargin}{0.5cm}

%\preprint{some number}

\title{The ZEPLIN--III Anti-Coincidence Veto Detector}% 

\author{D.Yu.~Akimov}\affiliation{Institute for Theoretical and Experimental Physics, Moscow, Russia}
\author{H.M.~Ara\'{u}jo}\affiliation{Blackett Laboratory, Imperial College London, UK}
\author{E.J.~Barnes}\affiliation{School of Physics \& Astronomy, University of Edinburgh, UK}
\author{V.A.~Belov} \affiliation{Institute for Theoretical and Experimental Physics, Moscow, Russia}
\author{A.A.~Burenkov} \affiliation{Institute for Theoretical and Experimental Physics, Moscow, Russia}
\author{V.~Chepel}\affiliation{LIP--Coimbra \& Department of Physics of the University of Coimbra, Portugal}
\author{A.~Currie}\affiliation{Blackett Laboratory, Imperial College London, UK}
\author{B.~Edwards}\affiliation{Particle Physics Department, Rutherford Appleton Laboratory, Chilton, UK}
\author{V.~Francis}\affiliation{Particle Physics Department, Rutherford Appleton Laboratory, Chilton, UK}
\author{C.~Ghag\footnote{corresponding author: c.ghag@ed.ac.uk}} \affiliation{School of Physics \& Astronomy, University of Edinburgh, UK}
\author{A.~Hollingsworth} \affiliation{School of Physics \& Astronomy, University of Edinburgh, UK}
\author{M.~Horn}  \affiliation{Blackett Laboratory, Imperial College London, UK}
\author{G.E.~Kalmus} \affiliation{Particle Physics Department, Rutherford Appleton Laboratory, Chilton, UK}
\author{A.S.~Kobyakin} \affiliation{Institute for Theoretical and Experimental Physics, Moscow, Russia}
\author{A.G.~Kovalenko} \affiliation{Institute for Theoretical and Experimental Physics, Moscow, Russia}
\author{V.N.~Lebedenko}  \affiliation{Blackett Laboratory, Imperial College London, UK}
\author{A.~Lindote} \affiliation{LIP--Coimbra \& Department of Physics of the University of Coimbra, Portugal}
\affiliation{Particle Physics Department, Rutherford Appleton Laboratory, Chilton, UK}
\author{M.I.~Lopes} \affiliation{LIP--Coimbra \& Department of Physics of the University of Coimbra, Portugal}
\author{R.~L\"{u}scher} \affiliation{Particle Physics Department, Rutherford Appleton Laboratory, Chilton, UK}
\author{K.~Lyons} \affiliation{Blackett Laboratory, Imperial College London, UK}
\author{P.~Majewski} \affiliation{Particle Physics Department, Rutherford Appleton Laboratory, Chilton, UK}
\author{A.St\,J.~Murphy} \affiliation{School of Physics \& Astronomy, University of Edinburgh, UK}
\author{F.~Neves} \affiliation{LIP--Coimbra \& Department of Physics of the University of Coimbra, Portugal}
\affiliation{Blackett Laboratory, Imperial College London, UK}
\author{S.M.~Paling}\affiliation{Particle Physics Department, Rutherford Appleton Laboratory, Chilton, UK}
\author{J.~Pinto da Cunha} \affiliation{LIP--Coimbra \& Department of Physics of the University of Coimbra, Portugal}
\author{R.~Preece} \affiliation{Particle Physics Department, Rutherford Appleton Laboratory, Chilton, UK}
\author{J.J.~Quenby}  \affiliation{Blackett Laboratory, Imperial College London, UK}
\author{L.~Reichhart} \affiliation{School of Physics \& Astronomy, University of Edinburgh, UK}
\author{P.R.~Scovell} \affiliation{School of Physics \& Astronomy, University of Edinburgh, UK}
\author{V.N.~Solovov} \affiliation{LIP--Coimbra \& Department of Physics of the University of Coimbra, Portugal}
\author{N.J.T.~Smith} \affiliation{Particle Physics Department, Rutherford Appleton Laboratory, Chilton, UK}
\author{P.F.~Smith} \affiliation{Particle Physics Department, Rutherford Appleton Laboratory, Chilton, UK}
\author{V.N.~Stekhanov} \affiliation{Institute for Theoretical and Experimental Physics, Moscow, Russia}
\author{T.J.~Sumner} \affiliation{Blackett Laboratory, Imperial College London, UK}
\author{R.~Taylor} \affiliation{School of Physics \& Astronomy, University of Edinburgh, UK}
\author{C.~Thorne} \affiliation{Blackett Laboratory, Imperial College London, UK}
\author{R.J.~Walker} \affiliation{Blackett Laboratory, Imperial College London, UK}

\date{\today}% It is always \today, today,
             %  but any date may be explicitly specified

\begin{abstract}
\noindent The design, optimisation and construction of an anti-coincidence veto detector to complement the ZEPLIN--III 
direct dark matter search instrument is described. 
One tonne of plastic scintillator is arranged into 52 bars individually read out by photomultipliers and 
coupled to a gadolinium-loaded passive polypropylene shield. 
Particular attention has been paid to radiological content. 
The overall aim has been to achieve a veto detector of low threshold and high efficiency without the creation 
of additional background in ZEPLIN--III, all at a reasonable cost. 
Extensive experimental measurements of the components have been made, including radioactivity levels and performance characteristics. 
These have been used to inform a complete end-to-end Monte Carlo simulation that has then been used to calculate the expected performance 
of the new instrument, both operating alone and as an anti-coincidence detector for ZEPLIN--III.
The veto device will be capable of rejecting over 65\% of coincident nuclear recoil events from neutron background in the energy range of interest in ZEPLIN--III.
This will reduce the background in ZEPLIN--III from $\simeq$0.4 to $\simeq$0.14 events per year in the WIMP acceptance region, a significant factor in the event of a non-zero observation.  
Furthermore, in addition to providing valuable diagnostic capabilities, the veto is capable of tagging over 15\% for $\gamma$-ray rejection, 
all whilst contributing no significant additional background. 
In conjunction with the replacement of the internal ZEPLIN--III photomultiplier array, 
the new veto is expected to improve significantly the sensitivity of the ZEPLIN--III instrument to dark matter, 
allowing spin independent WIMP-nucleon cross sections below 10$^{-8}$~pb to be probed.
\end{abstract}

% PACS, the Physics and Astronomy
                             % Classification Scheme.
%Use showkeys class option if keyword
                              %display desired

\maketitle

\section{INTRODUCTION}

\noindent ZEPLIN--III~\cite{1,z3sim} is a two-phase (liquid/gas) xenon detector developed to observe low energy 
nuclear recoils resulting from the elastic scattering of galactic 
weakly interacting massive particles (WIMPs)~\cite{3}. 
A key feature of ZEPLIN--III is the ability 
to discriminate between incident particle 
species for each event by recording both vacuum ultraviolet 
scintillation light and electroluminescence from ionisation of the xenon target~\cite{4,5}.  
The ratio in signal strength from these channels differs for 
electron and for nuclear recoil interactions, allowing the efficient rejection of
most background events. However, in the absence of an 
electromagnetic interaction, weak and strong elastic scattering events 
appear identical, making the rejection of single 
scatter background neutron events difficult~\cite{6,7,8}.  
Additionally, even with good discrimination, for large exposures a 
small fraction of electron-recoil events can be misidentified as nuclear recoils. 
With predicted WIMP event rates of less than 0.1 events/kg/day, 
and with energy depositions of the order of 10 keV, dark matter 
detectors such as ZEPLIN--III must reduce the 
number of background events, especially due to neutrons, by a very large factor.  
Typically this is achieved through the use of high purity components of low radionuclide content, 
hydrocarbon and lead shielding external to the detector to attenuate and moderate 
neutrons and $\gamma$-rays originating in the local environment, as well as 
the operation of the detector at a site with large rock overburden to reduce the cosmic muon flux.  
Having been manufactured from low background components such as oxygen free copper, 
shielded with 30~cm of high density polypropylene and lead, and situated at the Boulby Underground Laboratory, UK, 
the background event rate of ZEPLIN--III is already greatly reduced.  
Sensitivity to spin-independent WIMP-nucleon cross sections as low as 8.1$\times$10$^{-8}$~pb 
for a WIMP mass of 55 GeV/c$^2$ has been demonstrated~\cite{z3fsr}, as well as a spin-dependent WIMP-neutron cross section as low as 1.8$\times$10$^{-2}$~pb~\cite{z3fsrsd}.  
However, to improve the sensitivity of the device further, ZEPLIN--III is being retrofitted with an active veto detector.  
Since WIMPs are highly unlikely to interact within a detector system twice, any event doing so may be discarded.  Following this upgrade, the predicted combined background event
 rate in ZEPLIN--III translates into much less than one count within the WIMP acceptance region for a one year dataset.  
 The implementation of this veto system, in conjunction with new custom-built ultra-low 
background photomultiplier tubes (PMTs), will allow ZEPLIN--III to probe spin independent WIMP-nucleon cross sections below the 10$^{-8}$~pb level.

Experimental measurements have been made of the individual 
performance and radiological content of all components. 
These results have been fed into an end-to-end Monte Carlo simulation
of the ZEPLIN--III instrument~\cite{z3sim}, coupled with the entire veto detector array located in the Boulby mine, to evaluate the overall performance. 
An iterative process has been followed in which, with the geometry of the scintillator defined, alternative additional 
components and light collection efficiency were evaluated, informed by further experimental data where necessary.  

In the following sections we first 
describe the general design of the veto.  We then provide
details of the measurements of the radioactivity content 
of the components, the performance of components, and the optimisation
process.  Finally we provide results from the evaluation of the predicted overall performance of the veto. 

\section{Overall design}\label{overalldesign}

\begin{figure}[ht]
\includegraphics[width=8.6cm]{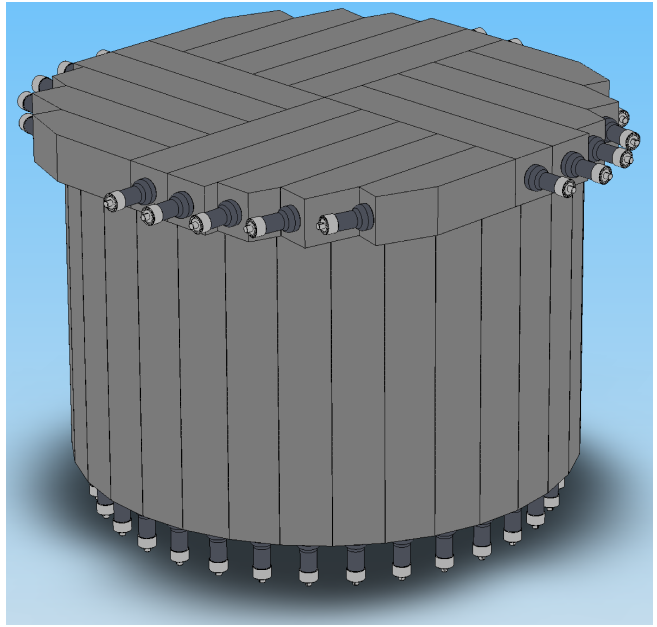}
\caption{\label{vetofig1}Three dimensional rendering of the ZEPLIN--III 
veto detector, showing only the plastic scintillator
bars with attached PMTs.  The diameter of the barrel is 160~cm}
\end{figure}

\begin{figure}[h]
\includegraphics[width=9.6cm]{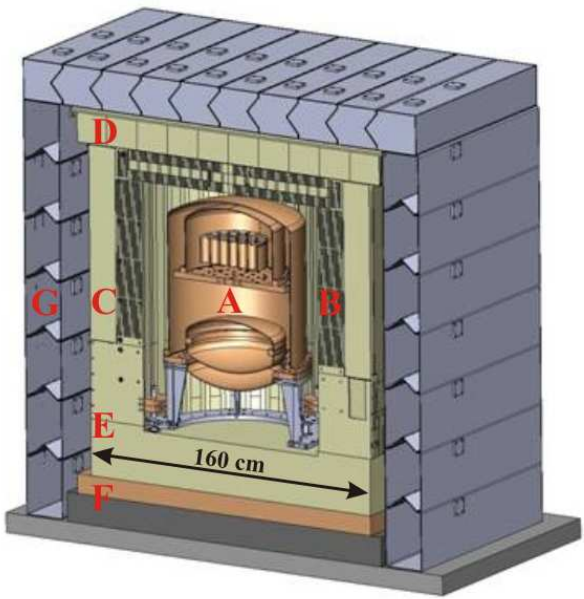}
\caption{\label{cad}
Cross-sectional view of the ZEPLIN--III experiment in its second science run configuration. 
In the centre is the ZEPLIN--III detector (labelled A) showing the copper vacuum vessel enclosing the xenon target and the LN2 vessel below it.  
Forming a barrel around ZEPLIN--III are the 32 Gd-loaded polypropylene pieces and roof plug (labelled B and hatched).  Surrounding these are the active
scintillator modules (C) with PMTs housed in cups and recessed into the lower polypropylene structure.  
The 20 roof modules of scintillator (D) rest on the roof plug.  
The lower polypropylene structure (E) contains no Gd and rests on a copper and a lead base (F).  
Finally, the lead castle (G) envelopes the entire assembly.}
\end{figure}

\noindent Figures~\ref{vetofig1} and~\ref{cad} depict schematically the veto system, while 
Table~\ref{table} presents key parameters of the overall system and its components.  It consists of 52 individual bars of  
plastic scintillator arranged so as to provide $>$3$\pi$~sr coverage around ZEPLIN--III. 
Thirty-two bars are 100~cm long, 15~cm thick parallelepipeds with trapezoidal cross section, that form a closed ring of 
outer diameter 160~cm (the `barrel'). These are arranged to stand on a continuous 30~cm thick base piece of passive 
polypropylene shielding.  A single 3-inch PMT is optically coupled to one end of 
each bar.  PMTs are a likely significant source of background radioactivity.  As such they are 
positioned on the lower face of the scintillator bars, 
in recessed holes in the base piece, thus keeping them 
farther from the ZEPLIN--III xenon target. 

\begin{table}\caption{\label{table}Key parameters of the ZEPLIN--III veto.}

\begin{tabular}{l c}
\hline
Total mass & 2323~kg \\
& \\
Plastic scintillator & UPS 923 A \\
Mass of plastic scintillator & 1057~kg \\
Number of bars & 52 \\
Typical Bulk Attenuation Length & 160~cm \\
 & \\
Mass of Gd loaded polypropylene & 756~kg\\
Slit pitch and width & 10~mm, 2~mm \\
Epoxy type & Rutherford Mix 71 \\
Volume of epoxy & 0.125~m$^{3}$ \\
Gd$_{2}$O$_{3}$ loading of epoxy & $>$2.7\% (w/w) \\
Average Gd loading of polypropylene & $>$0.5\% (w/w) \\
 & \\
Mass of polypropylene base & 510~kg \\ 
 & \\
PMT type & ETEL 9302KB\\
Number of PMTs & 52\\
Typical PMT QE at 420~nm & 27\% \\
 & \\
Data Acquisition & CAEN-1724 ADCs \\
Waveform digitisation & 100~ns \\
Waveform duration & -20 to +300~$\mu$s \\ 
 & \\
Contribution to ZEPLIN--III & $<$0.01 neutrons yr$^{-1}$\\
& $<$1000 $\gamma$-rays yr$^{-1}$ \\
Neutron tagging efficiency & $>$65\% \\
$\gamma$-ray tagging efficiency & $>$15\% \\
Mean neutron capture time & 35~$\mu$s \\
Effective threshold & $<$135~keV \\

\hline
\end{tabular}
\end{table}

Between the scintillator and the ZEPLIN--III instrument are positioned 32 pieces of 15~cm 
thick gadolinium (Gd) loaded polypropylene forming a barrel of outer diameter 130~cm. The scintillator 
and polypropylene barrels are rotated with respect to one another to ensure no line-of-sight gaps to the detector. 
Polypropylene has been chosen because of its
high hydrogen content, machinebility, and low radioactivity~\cite{ukdmweb}.
For ease of manufacture the gadolinium is added to the polypropylene in the form of 10~$\mu$m 
Gd$_{2}$O$_{3}$ powder mixed in to Rutherford Type 71 epoxy~\cite{epoxy}.  This mixture is set 
within 2~mm wide, 10~mm pitch vertical slots throughout the polypropylene.  
Monte Carlo simulations and experimental tests indicate that for neutron capture there is negligible difference 
between a uniform distribution and a discrete pattern, provided that the pitch of the slots does not exceed $\sim$15~mm 
(see section~\ref{pitchMC} below).  
The polypropylene and the scintillator 
parallelepiped bars are secured together to provide mechanical stability.  
Monte Carlo simulations have also been employed to confirm that mm-sized voids between or within attached standing scintillator modules result in neither a significant loss in neutron detection 
efficiency nor increased exposure to background for ZEPLIN--III (see section~\ref{voidsMC} below). 

A 15~cm thick, 115~cm diameter disc of gadolinium-loaded polypropylene forms a 
roof piece, with the remaining 20 plastic scintillator bars 
resting on it to form a roof that 
extends over the full diameter of the barrel modules, as shown in Figure~\ref{vetofig1}.  
A single PMT is optically coupled to the outer end of each roof bar, again maintaining maximum distance 
between veto PMTs and the xenon target.
The combined thickness of the passive and active hydrocarbon shielding is 30~cm, the same used in the first ZEPLIN--III science run~\cite{z3fsr}. 
The veto system is itself enclosed within the 
existing 20~cm thick Pb shielding of the 
ZEPLIN--III instrument (see Figure~\ref{cad}).

The geometry described above provides active rejection of background as follows. 
Firstly, neutrons entering the hydrocarbon shielding are efficiently moderated to 
thermal energies, mostly via H scattering, and undergo radiative capture 
predominantly on $^{157}$Gd (natural abundance 15.7\%) which has an extraordinarily 
high capture cross section of 2.4$\times$10$^{5}$ barns~\cite{gdcrosssec}. The level of gadolinium loading required was explored and is 
described in section~\ref{pitchMC}, where a value of at least 0.5$\%$ by weight, 
averaged over the polypropylene, is determined. The neutron capture is accompanied with the emission of typically 3--4 $\gamma$-rays 
of energy totalling $\sim$8~MeV, and at a loading of 0.5$\%$ these are emitted with a mean delay of 
35~$\mu$s from the time the neutron enters the loaded material. 
This is advantageous over a pure hydrocarbon system where radiative capture would occur after much longer periods of time and would lead to the release of a 
single 2.2~MeV $\gamma$-ray. Thus, with the gadolinium-loaded components located within the plastic scintillators, 
there is a very high probability of recording at least one significant energy deposition.  Such a design exposes the scintillator to 
an increased rate of neutron background from the environment relative to ZEPLIN--III, thereby reducing diagnostic capability, but greatly
increases the efficiency as an anti-coincidence detector for WIMP-like neutron recoils in ZEPLIN--III -- the primary function of the device.  
In the default mode of operation, data acquisition for the veto will be triggered predominantly by
energy deposition in ZEPLIN--III, with event-by-event veto waveforms recording for 20~$\mu$s before and 300~$\mu$s afterwards,
allowing for offline searching of coincidences.
Similarly, $\gamma$-ray emission, directly from components within ZEPLIN--III or otherwise, 
can lead to energy deposition in the scintillator.  
This allows coincident background events within ZEPLIN--III, predominantly from Compton scattering, to be identified.  In addition to operating as an active veto, this device will provide significant 
diagnostic capability for ZEPLIN--III, using independent triggering and providing a measurement of cosmic-ray muon fluxes, enhanced 
calibration capabilities and, by being a large volume low threshold device, permitting independent 
measurements of the neutron and $\gamma$-ray background environment in the laboratory around ZEPLIN--III.  

The veto design has several further advantages. Unlike with a liquid scintillator, where there is a possibility of 
leaks developing, the plastic veto scintillator bars are less of a chemical and fire hazard. Additionally, the modular design structure means that individual modules can easily be removed and 
repaired or modified if required without disturbing the remaining detectors or systems.  
As a diagnostic device, since signals generated in the veto are highly unlikely to appear 
in more than a few modules, a degree of directional reconstruction is possible, especially for cosmic-rays.  Thirdly,  
with the scintillator on the outside of the polypropylene, the veto has the ability to run in several modes simultaneously, with, for example, 
continuous recording of background as well as timelines where ZEPLIN--III has triggered, regardless of whether the veto has detected an event.  
Finally, the response of the entire system can be modelled accurately by incorporating the measured 
performance of each unit into a simulation, as opposed to the use of global parameters applied across a large volume and mass. 

\section{Radiological content of components}

\noindent At a vertical depth of 1070~m (2850~m water-equivalent shielding), the cosmic ray muon flux 
is reduced by a factor of $\sim$10$^6$ to a level of (3.79$\pm$0.15)$\times$10$^{-8}$ 
muons cm$^{-2}$ s$^{-1}$~\cite{muonn}. The resultant neutrons from cosmic ray muon spallation and secondary
cascades lead to a neutron scattering rate in ZEPLIN--III of $<$1 event/year for a nuclear recoil energy above 10~keV 
within the central 8~kg of xenon~\cite{muonal,noteal}. Thus, the most important background that must be considered comes from
the local environment of the laboratory and the instruments themselves. 
Consequently, all components proposed for use in the veto have been assayed for their radiological content.
The main contributions come from radioisotopes along the U and Th decay chains, which produce background neutrons via 
($\alpha$,n) reactions on other materials and from spontaneous fission, as well as numerous $\gamma$-rays.  The $\beta$-decay  
of $^{40}$K, which contaminates natural potassium at 100 ppm concentrations, generates 1.461~MeV $\gamma$-rays. To assess the overall neutron and $\gamma$-ray 
environment that the instruments will be exposed to, additional measurements of many other components 
have also been considered, for example the ZEPLIN--III instrument PMTs and the cavern rock. 
Two principal methods have been employed to determine the radioactivity content of components. 

\begin{table}\caption{\label{radtable}Radiological content with statistical uncertainties where appropriate of veto components as assayed either 
by direct observation of $\gamma$-ray emission (HPGe) or through mass spectroscopy techniques (ICP--MS/OES). See text for 
further details.}
\begin{tabular}{l c c c c}
\hline
Component & Mass, kg &   \multicolumn{3}{c}{Radiological content}  \\
          &       & U (ppb)    & Th 	(ppb)  	& K (ppm)  	\\
\hline
\\
{\em HPGe measurements} & & & & \\
Plastic scintillator	& 	1057.0		& 0.2$\pm$0.3 	& 0.1$\pm$0.7 	& 0.2$\pm$0.6	\\
PTFE inner wrap			&	8.9 		& 1.3$\pm$0.2 	& 0.2$\pm$0.5 	& 1.2$\pm$0.4 	\\	
Silicone				&	0.1 		& 2.9$\pm$0.4 	& 0.5$\pm$0.8 	& 5.7$\pm$1.1	\\
PTFE tape				& 	3.1  		& 3.2$\pm$1.3 	& 6.1$\pm$1.1	& 3.9$\pm$1.0	\\
Veto PMTs				&   6.2 	    & 38.0$\pm$0.8 	& 21.1$\pm$1.2 	& 65.5$\pm$2.4 	\\
PMT preamplifiers		&  	0.7         & 8.4$\pm$1.7  	& 13.2$\pm$2.2 	& 10.1$\pm$1.7	\\
PMT base				& 	5.5 		& 12.7$\pm$1.4 	& 14.8$\pm$2.4 	& 20.2$\pm$2.4 	\\  
Epoxy					& 	70.0		& 2.5$\pm$0.6	& 0.9$\pm$0.3	& 0.6$\pm$0.1 	\\
Gd oxide				& 	8.0			& 0.9$\pm$0.1 	& 1.2$\pm$0.3 	& 1.7$\pm$1.1 	\\  
\\
{\em ICP--MS/OES}    & & & & \\

Copper tape				& 	26.0		& 1.9$\pm$0.2 	& 2.9$\pm$0.4 	& 14.0$\pm$2.0 	\\
PTFE inner wrap			& 	8.8 		& 2.0$\pm$1.0 	& 5.0$\pm$1.0 	& $<$4		\\
Veto PMTs				&   6.2 		& 30.2$\pm$2.2	& 30.0$\pm$3.7	& 60$\pm$2.2	\\ 
PMT preamplifiers		&  	0.7 		& 10.3$\pm$0.5	& 29.7$\pm$3.2	& 24$\pm$3.7	\\ 
PMT base				&	5.5 		& 13$\pm$3.4	& 19$\pm$2.0	& 21$\pm$3.0	\\ 
Polypropylene			&  	510 		& $<$1		& $<$1		& $<$5		\\
PMT mounting			&  	15.8 		& 30$\pm$7.8	& $<$10		& $<$10		\\
Cabling					& 	30.2 		& 110$\pm$5.4	& 20$\pm$3.2	& 29$\pm$7.3	\\
Connectors				& 	2.1 	   	& $<$10		& $<$10		& $<$4		\\
Optical gel				&  	0.3 	   	& $<$1		& $<$1		& $<$1		\\
Gd oxide				& 	8.0			& 2.5$\pm$0.5 	& 3.4$\pm$0.7 	& $<$4		\\ 
\hline
\end{tabular}
\end{table}

\subsection{\label{gemeasurements}Direct $\gamma$-ray measurements}
\begin{figure}[ht]
\includegraphics[width=9.5cm]{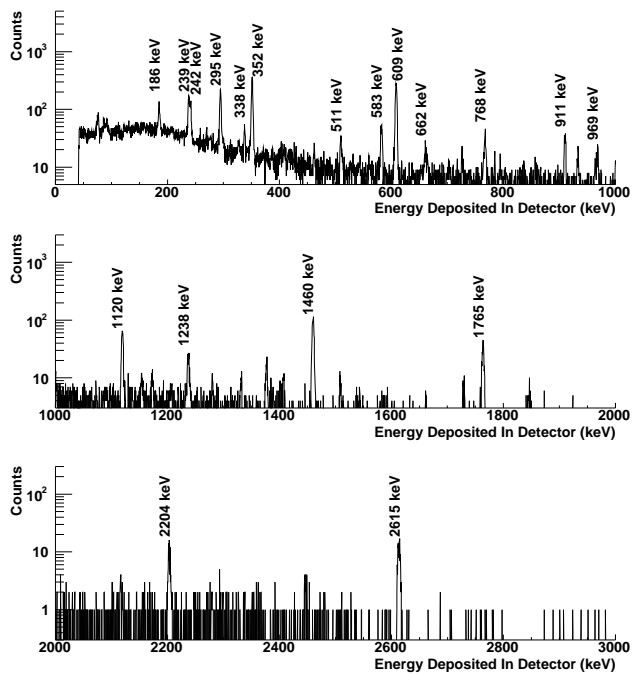}
%\vspace*{5cm}
\caption{\label{Gespec} Illustrative spectra from the underground HPGe detector. In this case
the sample being measured was a collection of three PMTs, positioned close to the HPGe detector head. Shown is 
a background-subtracted exposure of 160~hours duration.  The energies of $\gamma$-ray lines of interest are indicated.} 
\end{figure}

\noindent Direct measurements of the $\gamma$-ray emission from candidate components has been made using 
a high purity germanium (HPGe) detector located in a dedicated low background counting facility at the Boulby mine.  
The HPGe detector head was encased in an inner copper and outer lead castle leaving a (30~cm)$^{3}$ test cell.  
The ambient $\gamma$-ray flux experienced by the HPGe detector in the absence of a test sample was then 
assessed through a week-long background run. 
A sample of material was placed close to the head and further data accrued over a period of several days. 
An example for a background-subtracted sample spectrum is shown in Figure~\ref{Gespec}, revealing 
photopeaks at energies corresponding to known $\gamma$-ray emissions from U and Th chains and from K as well as other dominant lines of interest. 
To assess the quantity of an isotope that must have been present in order to generate the measured excess
in the background-subtracted spectrum, secular equilibrium was assumed and a GEANT4~\cite{geant4} Monte Carlo simulation of the low background counting setup performed.  Each simulation was individually tailored
to the geometry of the sample being tested.  
Sensitivities at the level of 0.1 ppb for U and Th, and 0.1 ppm for K were achieved using this technique.
The conversion between $\gamma$-rays emitted from an isotope and the source contamination level is given 
using the following $\gamma$-ray yields per kg$\cdot$day: 2310 $\gamma$-rays/ppb U, 958 $\gamma$-rays/ppb Th and 285 $\gamma$-rays/ppm K.  
A summary of the activity levels found is presented in Table~\ref{radtable}, together with the total mass
of each component used in the construction.

\subsection{Mass spectrometry measurements}
\noindent To supplement some of the $\gamma$-ray measurements, and to provide a cross check of their accuracy, 
a subset of components was measured using different techniques at the Geosciences Advisory Unit (GAU--Radioanalytical) at the 
University of Southampton.  Samples were dissolved and subsequently diluted in 2\% HNO$_{3}$. Activity due to $^{232}$Th, $^{235}$U and $^{238}$U was 
measured using an inductively-coupled plasma mass spectrometer (ICP--MS), 
while potassium content was assessed using an inductively-coupled plasma optical emission spectrometer (ICP--OES). A calibration 
$^{236}$U recovery tracer was added. The results from this process are also presented in Table~\ref{radtable}.  
Differences arise due to the assumption of secular equilibrium used to infer contamination levels
in the direct $\gamma$-ray measurements, as opposed to the direct measurement of atoms using mass-spectroscopy.  In all cases, upper limits from either set of 
measurements (HPGe or ICP--MS/OES) have been used as inputs for simulations, described in Section~\ref{MC}, performed to assess the impact of radiological contamination in the veto components on
ZEPLIN--III as well as in the veto itself.

\section{Performance of Components}
\noindent The overall capability of the veto relies on the performance of individual components as well as on the 
geometry and methodology in which they are used. In this section we report on the technical performance of 
key components: the PMTs, the plastic scintillator, the front-end electronics and the data acquisition system.
 
\subsection{Photomultipliers}\label{pmts}

\noindent The PMTs used are model 9302KB from ET Enterprises Ltd (ETEL).  
These are 78~mm diameter window devices with bialkali photocathode and 9 
dynode stages.  Their envelopes are manufactured from low background glass, 
with manufacturer-quoted contamination levels of 30 ppb U, 30 ppb Th and 60 ppm K.  
As described in Section~\ref{gemeasurements}, 
experimentally measured values using a HPGe detector in a 
low background environment confirm these levels.% as upper limits.

Voltage divider networks (ETL C647BFN2-01) 
were fitted to low background B14A bases supplied by ETEL, with U, Th and K contamination levels 
measured to be consistent with, or lower than, the manufacturer stated values of 13~ppb, 19~ppb and 21~ppm, respectively.  
To amplify the small signals from low energy events and to impedance-match the output of the PMT to the 
data acquisition digitisers, pre-amplifiers have been designed and manufactured that give a gain 
of $\sim$10.  
The preamplifiers are tooled onto surface mount printed circuit boards made 
from light-weight low-background single-sided resin bonded paper.  
The PMTs and their bases (including the voltage divider network and the preamplifier board) are secured within 
a PVC cylinder which is itself chemically bonded directly to the scintillator. 
Springs mounted in the cylinder cap provide support for the PMT, pushing it into contact
with the scintillator surface via BC-630 optical coupling grease. 
The PMTs are operated with negative bias supplied by Lecroy 1443NF boards within a Lecroy 1440 HV high voltage system.
All power supply and signal cabling for the photocathode-anode (k-a) HV, pre-amplifier power, and signal output, 
has been manufactured to order with minimal radiological contamination.

\begin{figure}[ht]
\includegraphics[width=8.6cm]{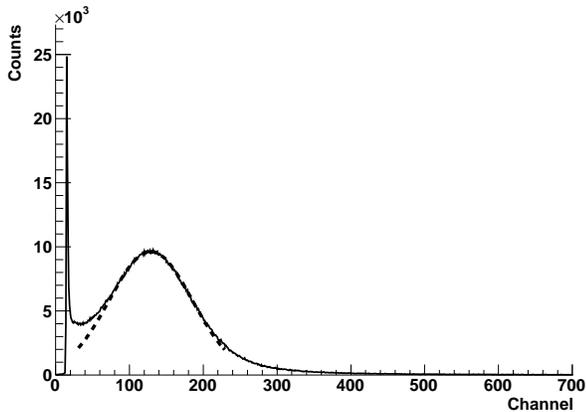}
\caption{\label{spespec} A typical single photoelectron spectrum (solid line), with a partial Gaussian fit to the peak (dashed line), obtained from a PMT in the 
absence of a light source and with no plastic scintillator attached.}
\end{figure}

Single photoelectron (SPE) integrated pulse height spectra from dark thermionic emission have been obtained for each PMT, an example of which is shown in Figure~\ref{spespec}.  
The average dark count rate is found to be approximately 300~Hz.  
By design, the 10~MHz sampling frequency of the data acquisition is sufficient for 
good recovery of SPE signals, allowing these to be used routinely for relative gain characterisation.  
These spectra were taken with no scintillator or other light sources present. 
For these measurements a single reference PMT was used assuring environmental and electrical uniformity across tests, 
with both the reference and test PMT housed in a light-tight box. 
The bias voltage of the test PMT was varied until the spectral peak position observed from single 
photoelectron emission from the photocathode was matched to that of the reference PMT.  This was repeated for all PMTs.

With PMT bias voltages adjusted to yield normalised SPE spectra, the quantum efficiency (QE) and gain of each PMT was calibrated. 
This was performed by illuminating the test PMT photocathode with a blue LED to stimulate emission from the photocathode.  The number of photoelectrons observed was 
then calculated using two different methods.  

In the first method, a statistical approach is adopted where the number of photoelectrons detected is determined from the width of the distribution generated. 
The width of the peak seen from the LED is dominated by the Poissonian statistics of the photoelectron emission, 
{\it i.e.} $\sigma = m/\sqrt{N}$ where $\sigma$ is the 
measured standard deviation, $m$ is the centroid of the peak, 
and $N$ is the mean number of photoelectrons emitted for events in the peak. 
This will underestimate the total number of photoelectrons from a PMT, since it does not account for additional contributions to
$\sigma$, the most important being the width of the SPE spectrum itself.  However, as was the case here, this can be mitigated somewhat by generating a large number of photoelectrons such that the 
width of the distribution is much greater than the width of the SPE spectrum. 

\begin{figure}[ht]
\includegraphics[width=8.6cm]{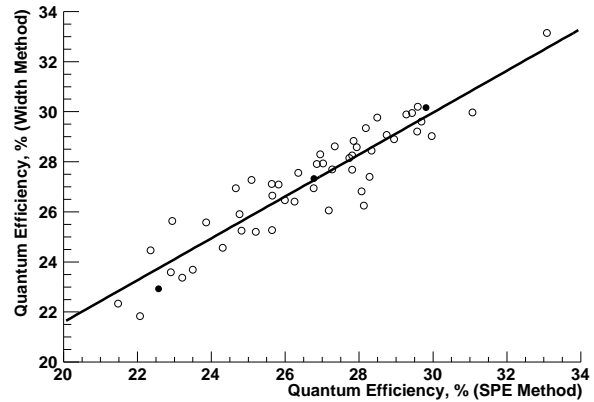}
\caption{\label{qe}The quantum efficiencies of the PMTs used in the ZEPLIN--III veto.  Two methods have been used
giving broadly consistent results.  Three PMTs spanning the range of performance were assessed for absolute 
quantum efficiency externally; these are indicated by the filled symbols.  A linear fit through the data points is drawn to
highlight the bias between the two methods used.}
\end{figure}

The second and more traditional method is to calculate the number of detected photoelectrons by simply   
dividing the peak position of the LED spectrum by that obtained from the single photoelectron response from dark spontaneous emission.  
The measured response was then normalised to absolute QE when 3 out of the 52 PMTs were absolutely calibrated at ETEL.  
The distribution of QEs, as plotted in Figure~\ref{qe}, shows good agreement between both methods used for estimating the number
of photoelectrons.  The gain normalised measurements confirmed the relative QEs between PMTs for sensitivities of 50~A/lm and 200~A/lm 
as quoted by the manufacturer.  
The mean QE at 360~nm is 30.1\% and at 420~nm is 27.1\%.

\subsection{Plastic Scintillator: UPS--923A}

\noindent  The 1057~kg of scintillator material is polystyrene based UPS--923A (p-terphenyl 2\%, POPOP 0.02\%) 
produced by Amcrys--H, Kharkov, Ukraine~\cite{Amcrys-Hwebsite}.  
This material emits scintillation light with a peak intensity at 420~nm and 
with a rise time of 0.9~ns and decay time of 3.3~ns.  
The density of UPS--923A is 1.06 g~cm$^{-3}$ and its refractive index is 1.52.  
The nominal light output for the material is stated as 55\% of anthracene.  
The transparency of the scintillator material is determined by the so-called Bulk Attenuation Length 
(BAL) which is the length that reduces the initial light intensity by factor e according to the Buger–-Lambert law, 
and is expected to be greater than 1~m (see later for measured values).
A previous study by the manufacturers suggests 
this material to have long-term stability, such that there is no natural 
degradation reported over a 12~year period~\cite{D13-2005-1111 UPS923A}.  

\begin{figure}[ht]
\includegraphics[width=8.6cm]{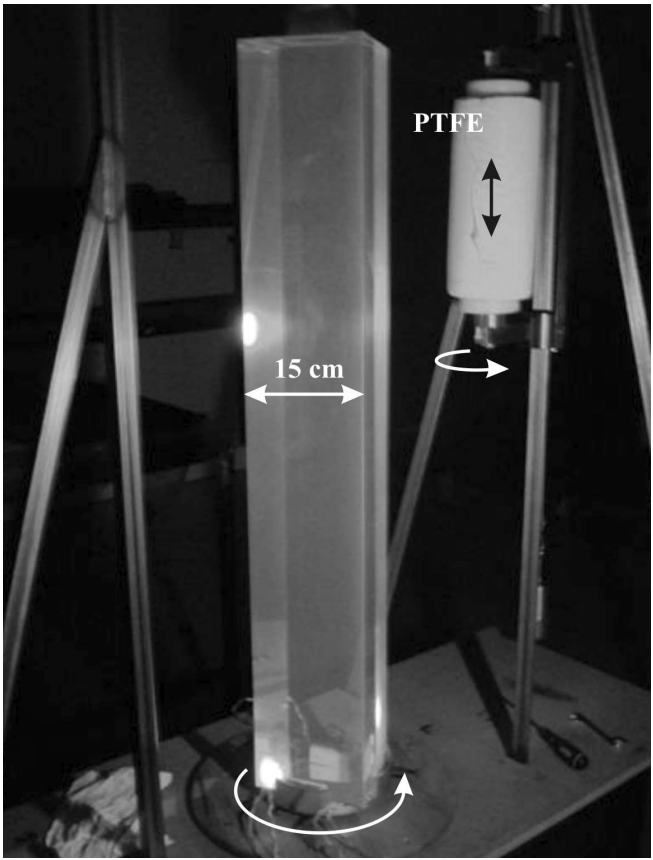}
\caption{\label{unwrapped} Photograph of an unwrapped barrel bar of plastic scintillator with the apparatus used to wrap bars with PTFE to increase light collection.} 
\end{figure}

\begin{figure}[ht]
\includegraphics[width=8.6cm]{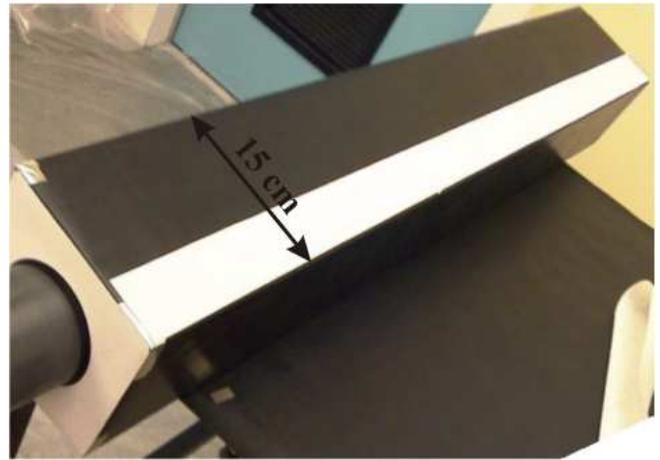}
\caption{\label{wrapped} Photograph of a wrapped plastic scintillator module.  Around the bare plastic is a layer of PTFE, a mirror at the opposite end from the PMT, and black
wrapping to light tighten the module.  The PMT itself is housed in a light tight cup bonded to the scintillator.} 
\end{figure}

To minimise loss of light within a scintillator bar each 
piece has been wrapped along its length with PTFE sheet giving it a high diffuse reflectivity.  
Experimental tests were performed comparing a number of diffuse and specular reflective 
wrapping materials including white paper, aluminised Mylar, 
Vituki~3M~\cite{vituki} and other PTFE-based sheeting.  
Three layers of 76~$\mu \rm m$ thick PTFE film~\cite{innerwrap} 
was found to provide the most cost effective option, 
increasing light collection by up to a factor of 2 at the PMT.  At the far end from the PMT a specular reflective 
aluminised Mylar foil is placed that acts as a mirror to improve the uniformity and increase the 
effective BAL of the unit.  The PMT is optically coupled to the scintillator with BC--630
silicone-based optical couplant and is housed in a PVC cylindrical tube chemically 
bonded to the scintillator.  
The entire assembly has then been wrapped in black opaque PVC sheet~\cite{outerwrap} to provide light tightness.
Figures~\ref{unwrapped} and~\ref{wrapped} show unwrapped and wrapped plastic scintillator barrel bars.

The performance of each of the 52 pieces has been assessed individually as follows. 
For each fully wrapped plastic scintillator module, spectra of the PMT response 
to a $^{22}$Na calibration source were recorded for each of six positions along the length.  To ensure consistency, the same PMT, 
voltage divider network, preamplifier and operating bias
were used in these measurements.  
Additionally, the use of optical grease was omitted to reproduce conditions as exactly as possible across scintillators.  
It was found that without the optical grease the number of photoelectrons detected is approximately halved, and this is confirmed by 
Monte Carlo simulations~\cite{EJBthesis}.
The measurements were performed in a surface laboratory of predominantly concrete construction, 
and thus, for plastic scintillators of this size,
a high rate of $\gamma$-rays from the local environment was observed. 
To reduce this contribution, events 
seen in the plastic scintillators were required to be in coincidence with events in a 
3" diameter, 3" long NaI scintillation detector placed on the other side of the 
radioactive source from the plastic scintillator. Since $^{22}$Na decays by $\beta^{+}$ emission 
this greatly enhanced the fraction of events caused by the 511~keV annihilation and 1275~keV 
de-excitation $\gamma$-rays.  
However, since $\gamma$-rays produce largely Compton electrons in organic scintillators at these energies,
the trigger signal of the NaI detector can come either from another $\gamma$-ray, or from the Compton-scattered
$\gamma$-rays detected in the plastic scintillator.  In the latter, the 
shape of the energy spectrum may be effected since Compton electrons in the scintillator deposit lower energy and yet the scattered $\gamma$-ray 
retains sufficient energy to trigger the NaI.  
Monte Carlo simulations of the experimental setup confirmed that, with the geometry used for these measurements, the rate of such events relative 
to true coincidence triggers from separate gamma-rays was negligible.

\begin{figure}[ht]
\includegraphics[width=8.6cm]{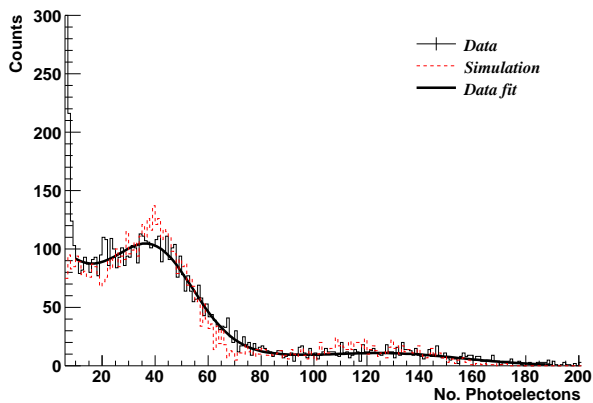}
%\vspace*{6cm}
\caption{\label{balfig} Typical spectrum (solid histogram) showing the response of a single barrel module of 
plastic scintillator to irradiation with a $^{22}$Na $\gamma$-ray source. See text for further details 
of the source geometry. The solid line is the result of a fit to the data assuming partial Gaussians for the peaks 
attributed to the 511 and 1275~keV $\gamma$-rays and an exponential to describe the background.
Also shown (dashed histogram) is the result of a Monte Carlo simulation of this exposure.
} 
\end{figure}

A typical Compton spectrum is shown in Fig.~\ref{balfig}, revealing 
features corresponding to the 511~keV $\gamma$-rays, as well as a weaker peak 
corresponding to the 1275~keV $\gamma$-ray emission following $\beta^{+}$ decay of $^{22}$Na, and a background
well described by an exponential trend. 
The calibration between observed Compton peak position and number of photoelectrons has been used to
allow this spectrum to be plotted in terms of numbers of photoelectrons. 
A two-partial-Gaussian-plus-exponential fit was made to the data and the movement of the
centroid of the partial Gaussians and the Compton edge as a function of the position of the source along the length was 
used as the indicator of the technical attenuation length (TAL). 
The BAL describes the attenuation of photons in a 
beam due to the scintillator only, whereas the TAL parameter is an auxiliary parameter. The difference between the and BAL comes
from the fact that the real light path from the vertex of the interaction to the PMT is longer than a straight line path, since most of
of the light reaching the PMT has been reflected at the surface of the scintillator bar through total internal reflection.  
The measured TAL combines geometry effects (light path, reflection),
light attenuation, wavelength shift (due to wavelength dependent absorption and re-emission) and
wavelength dependent quantum efficiency of the photo cathode.  
In the case of the 
modules measured here, a highly-reflective specular mirror surface has also been placed at the opposite end to 
the PMT. Consequently, one expects, to first order, the response as a function of distance from the
PMT to vary as the sum of two exponentials, one describing the reduction due to the separation of  
PMT and source position, and one which describes the reduction due to the distance between the PMT and the 
image of the source in the mirror:  
$$
S(x)=Ae^{-x/T}+Ae^{-(2l-x)/T}
$$
where $S(x)$ is the centroid of the measured photopeak, $x$ is the distance from the PMT face, $l$ is the
length of the scintillator module (the distance between the PMT and the mirror), $T$ is the TAL and 
$A$ an arbitrary scaling constant. In principle, one could use a different scaling for the
term describing the response due to reflections from the mirror, effectively accounting for 
imperfect reflection from it, but the quality of the data available did not justify an additional free parameter.
The measurements of TAL have been complemented with cosmic ray muon measurements in 
which the NaI coincidence detector was replaced by two small plastic scintillator detectors 
placed above and below the test piece. In this case the spectra obtained are well 
described and fitted by a Landau distribution, and again the 
response as a function of measurement position determined.
Consistent results were found for the TAL measured either with the $^{22}$Na source or with cosmic rays. 

\begin{figure}[ht]
\includegraphics[width=8.6cm]{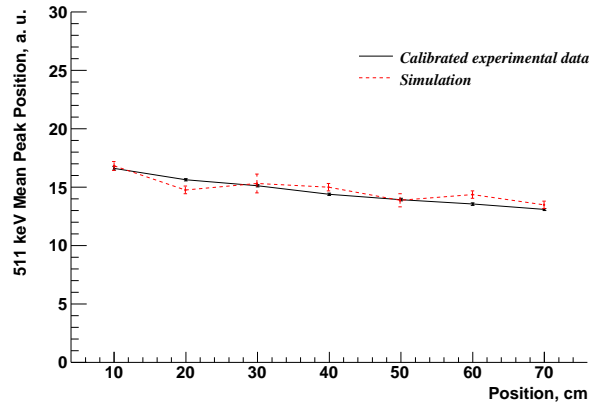}
\caption{\label{responsevspos} Variation of the peak position from the 511~keV $\gamma$-ray with source location both for data from one of the 1~m long barrel modules and 
evaluated from Monte Carlo simulations of those tests.  Lines have been drawn to guide the eye.} 
\end{figure}

\begin{figure}[ht]
\includegraphics[width=8.6cm]{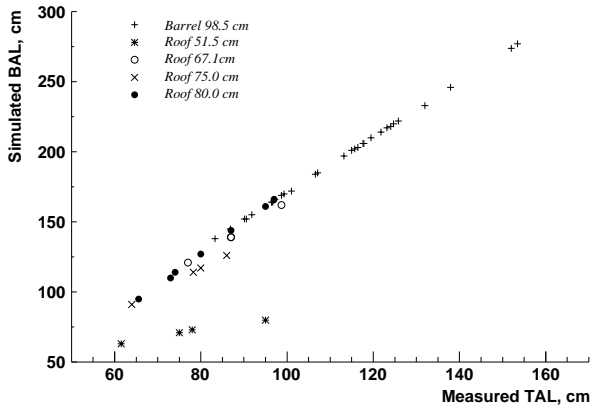}
%\vspace*{5cm}
\caption{\label{baltal} Measured technical attenuation length (TAL) and Monte Carlo predicted bulk attenuation length (BAL) for each bar of plastic scintillator.}
\end{figure}

To assess whether the measured TALs are consistent with the expected BAL of the material,
GEANT4 Monte Carlo simulations of these tests have been performed (see Section~\ref{MC}). These included the geometry of the plastic scintillator, 
laboratory and source, as well as physical parameters such as the light output of the plastic, gain and
quantum efficiency of the PMT (see section~\ref{pmts}). 
The simulations include full tracking of scintillation photons from $\gamma$-ray energy deposition
through to incidence on PMT photocathode and stochastic generation of photoelectrons, 
allowing realistic spectra to be generated. 
The lighter histogram in Figure~\ref{balfig} shows the result of the simulation tailored to 
that particular measurement's geometry, illustrating that the key features are very well reproduced. 
The simulations are repeated for different values of BAL, each time generating spectra
for different source target positions, and these spectra evaluated in the same manner
as was done for real data to extract a TAL. 
Figure~\ref{responsevspos} shows the position of the centroid of 
the peak attributed to the 511~keV $\gamma$-ray as a function of the source location both for real data from
one of the barrel modules, and evaluated from Monte Carlo simulations of those exposures with an 
appropriately chosen value of BAL. This BAL is then taken as corresponding to the 
experimental value of TAL indicted by the data.  
Lines have been added to guide the eye in this figure and in subsequent figures where appropriate.  
As expected, for each scintillator bar geometry a monotonic dependence between BAL and TAL is found, 
with TALs having smaller values than BALs, and the most uniform geometry bars having the highest TALs. 
These results are summarised in Figure~\ref{baltal}. We find a large dispersion in 
BAL, ranging from $\sim$60~cm to greater than 2.5~m, with significant variations 
even between bars polymerised in the same batch. 
The measurements allow the matching of individual bars of plastic scintillator 
to particular PMTs, matching lower BAL with higher QE to provide a more uniform performance across the 
entire veto and informing the placement of scintillator modules within the array. 
This also provides input for the Monte Carlo 
simulations described in section~\ref{ubersim} in which the
overall performance of the whole veto system is evaluated.

Although many processes contribute to the scintillation light emission~\cite{birks, matveeva}, the absolute light output of the plastic scintillator can be calculated by matching the measured and simulated spectra.  
A consistent value of $\sim$5500~ph/MeV was 
required for all bars, regardless of which polymerisation batch they came from. This is lower than expected, as 55\% of anthracene would correspond to nearer 8500~ph/MeV, necessitating the use of reflective wrapping around the scintillators.

During normal operation of the veto a blue LED unit will continuously flash 52 separate optic fibres coupled directly to the ends of the plastic scintillator modules furthest from the PMT.  
This will allow any degradation of the scintillator performance over time to be monitored.  

\subsection{Data Acquisition}

\begin{figure}[ht]
% \vspace*{5cm}
\includegraphics[width=8.6cm]{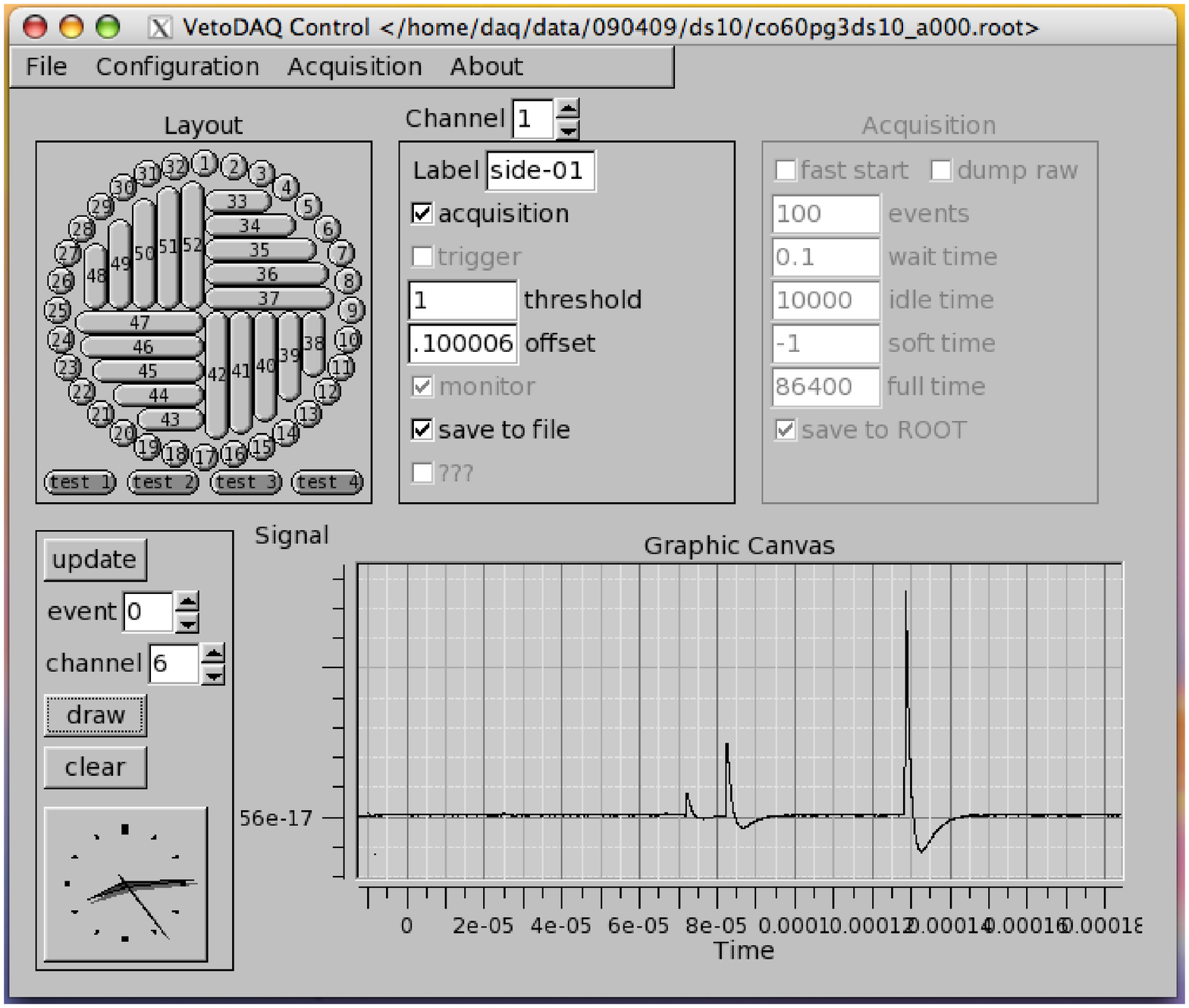}
\caption{\label{screenshot}Graphical user interface of veto data acquisition.}
\end{figure}

\noindent The outputs from the PMT preamplifiers are fed to CAEN V1724 ADCs housed in a VME8011 crate with CAEN V2718 
PCI bridge communicating via optical 
link to a dedicated Linux-based data acquisition computer.  The digitisers have $\pm$2.25~V input range with 14-bit resolution and 40~MHz bandwidth with a sampling rate of up to 100~MS/s 
simultaneously on each channel, with internal buffering to ensure zero deadtime 
performance at expected data rates. Custom made run-control software provides a graphical user interface for beginning and terminating data acquisition runs 
with easily changeable run parameters and toggles over various operation modes.  The software also provides remote capabilities to fully automate instrument operation.
A screenshot of this interface is shown in Figure~\ref{screenshot}.
Amongst the most important operation modes are the veto `slave' mode and calibration/diagnostic `master' mode.  

In {\em slave} mode, all modules take an external trigger sourced from the ZEPLIN--III instrument.  
As such, the veto is guaranteed to record data when ZEPLIN--III itself has recorded an event.  
Besides simplicity of design, this minimises data volume and obviates the need for a separate hardware-enforced trigger 
and associated efficiency loss at low energies.  Additionally, spontaneous photoelectron
emission from the PMTs, present in the timelines recorded where pulses are absent, can be used for equalising gains,
 calibrating and monitoring stability of the PMTs without acquiring dedicated data, where conditions between data runs may vary.

Finally, since even with high energy depositions in the 
veto only a few of the scintillator modules are involved, 
the remaining modules provide diagnostic information at the time of the event.
Monte Carlo simulations indicate background $\gamma$-ray events depositing 
$>$200 keV distribute $>$90\%  of the energy 
in 6 modules or fewer (see section~\ref{ubersim}).  

The expected time delay between signals in ZEPLIN--III and in the veto defines the requirements 
for the duration of the waveforms to be be recorded. Coincident $\gamma$-ray events
will be essentially simultaneous. For single scatter neutron 
recoils within ZEPLIN--III, where the veto signal is provided by energy deposition of $\gamma$-rays 
following neutron capture on gadolinium, the coincidence is expected to have
a mean delay of $\sim$35~$\mu$s (see Section~\ref{gdloading}). However, for small
energy deposits in ZEPLIN--III, the trigger is in fact provided by the S2 electroluminescence, which
is itself delayed by up to 17~$\mu$s (due to the time it takes electrons to drift through the xenon). 
It is therefore necessary that a period of pre-trigger data is also recorded:
values of 20~$\mu$s pre-trigger and 300~$\mu$s post-trigger have been chosen. 

\begin{figure}[ht]
\includegraphics[width=9.6cm]{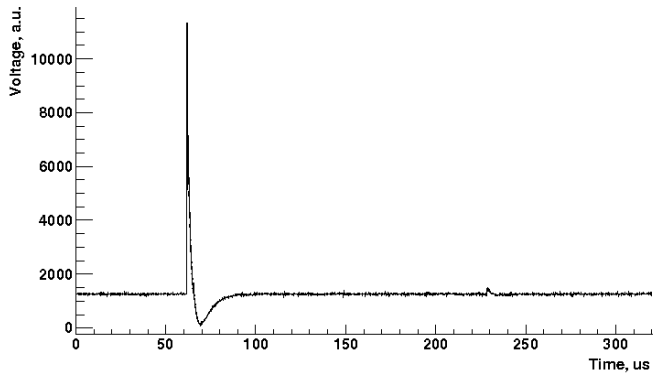}
\caption{\label{waveforms} An example waveform from a barrel module plastic scintillator in response to a 
$\gamma$-ray from a $^{60}$Co source, with the pulse starting at 60~$\mu$s.  Also clearly visible is a single photoelectron at 230~$\mu$s.}
\end{figure}

An example of a waveform recorded in the veto data acquisition is presented in Figure~\ref{waveforms}. 
This event corresponded to irradiation by a $^{60}$Co source 
located approximately half way along the length of 
one of the barrel modules. By design, the gain of the system is such that $\gamma$-rays of about this energy
populate the mid range of the 14-bit ADC. The baseline 
noise is observed to be between typically 2~mV, or about 30 ADC channels.  In addition to the 
$\gamma$-ray event at 60~$\mu$s, a typical single photoelectron can be seen at 230~$\mu$s, clearly visible above the noise level. 

The {\em master} diagnostic/calibration mode takes either a forced trigger to allow the veto modules to record effectively continuously without imposing a signal threshold for triggering, or can be set to record data
 when a defined number of scintillator modules detect a signal above a defined threshold.  These are necessary for 
{\em in situ} calibration of the veto device and for recording background data in all modules.  As such, when operating
in this master mode, the veto system can operate as a detector independently of ZEPLIN--III if required.

The default operating mode will be a combination of these triggering techniques, concurrently recording a level of background without hindering the coincidence detection capabilities: 
the veto will trigger whenever ZEPLIN--III observes a signal, regardless of the presence of 
a signal in the veto, but it will also trigger whenever the veto observes a signal in two or more modules above some threshold set to exclude noise
and maintain a low count rate.  Since the acquisition can run with zero deadtime by storing events in its internal buffer, there is no possibility
of missing a ZEPLIN--III trigger.  To maintain accurate synchronisation between ZEPLIN--III and the veto data acquisitions, a dedicated hardware unit 
produces a digital time stamp which is sent to both acquisitions so that events can be correlated with 100~ns accuracy.  

Finally, taking advantage of the segmented nature of the veto, PMT outputs from the 20 roof modules are fed into a dedicated triggering unit that passes signals through to 
the ADC inputs but also sums them and shapes this summed signal.  
The shaping parameters are set such that if the unit finds a pulse in any module that is characteristic of a cosmic ray muon, all 52 veto channels are triggered and the event recorded,  
regardless of a coincident event in ZEPLIN--III.  
This is a valuable diagnostic allowing measurement of the cosmic ray
flux through the laboratory, as well as providing a measure of muon induced neutron background in the local environment.  

\section{Monte Carlo simulations}\label{MC}

\noindent Monte Carlo simulations have been used extensively in this work, both to
characterise individual components, for example correlating the BAL and TAL responses 
of individual plastic scintillator modules, and as a tool for optimising and 
understanding the overall performance of the veto. 

The requirements for this Monte Carlo simulation are relatively demanding.  
The programs should be capable of modelling the transport and 
interactions of particles (essentially neutrons and $\gamma$-rays) internal and external to 
ZEPLIN--III and the veto through to the production of electron and nuclear recoils.  
The program must then simulate the physical processes involved in the generation of the optical 
response to scintillation (and electroluminescence in ZEPLIN--III), and finally generate 
the response for all channels in order to produce realistic datasets.  
The existing ZEPLIN--III Monte Carlo simulation package~\cite{z3sim}, and the veto simulations developed here, 
have utilised the GEANT4 toolkit~\cite{geant4}, which is capable of fulfilling all of these requirements.  

In this section we first present the general features of the simulation, leading to calculation of 
the expected ability of the veto to reject both neutron and $\gamma$-ray events that, based on 
ZEPLIN--III data alone, might be misidentified as WIMPs. A key feature of the veto design is 
the inclusion of gadolinium to enhance the response to neutrons. We have used the Monte Carlo methodology to
explore the dependence of the design to the amount and location of gadolinium used, and this is presented.
We then bring together the measurements of radiological content and
performance of all the veto components to estimate the absolute event rate that the veto will experience,
and its impact on the rejection performance of the instrument.
Finally, we utilise the Monte Carlo simulations to explore effects such as misalignment between modules introduced during assembly.  
The Monte Carlo simulation has also been used to inform numerous other choices not mentioned, 
for example the geometry of the active and passive modules.

\subsection{General features of the simulation}

\noindent The GEANT4--v9.1 Monte Carlo toolkit has been used. 
All simulations included accurate geometry of the plastic scintillators, their wrapping including
the 95\% specular reflective mirror at the far end, 
the PMTs including envelope, photocathode, internal elements and PVC support structure and
optical grease (in terms of changes to refractive indices).
Also incorporated are the geometry and materials of the surrounding area; for simulations of the
whole veto this includes an extensive description of the Boulby laboratory with 
full details of ZEPLIN--III itself, the Gd and all related processes, the local polypropylene and lead shielding, 
the building, and rock walls. By varying the depth of rock included in the simulation it was found
that a depth of 3~m for neutrons and 0.25~m for $\gamma$-rays is sufficient to accurately simulate the flux emanating from these surfaces 
($<$1\% contribution to uncertainty). 
For simulations of measurements made in surface laboratories, the details of
the walls, tables and other large objects in the vicinity were included as necessary. 
The light yield and BAL of the scintillator and the single photoelectron response 
and the QE of the PMTs are taken from measurements described earlier.

Simulations have been performed for radiation 
originating from the cavern walls, from ZEPLIN--III components, from the veto components
and from calibration sources and, in each case, measured values of activity  
have been used. In the case of U and Th, the SOURCES--4C~\cite{sources} 
code has been used to generate the emission spectra
for neutrons. All particles and their secondaries are tracked.
The location, size and event time of all energy depositions are recorded.  

A nuclear recoil energy deposition generates much less electronic excitation 
(primary scintillation and ionisation) than an electron recoil event of similar energy due to `quenching'.  A quenching coefficient is defined as
the ratio of the amount of light induced by a recoil nucleus to the amount of light induced by an electron of the same kinetic energy.  
A (conservative) quenching coefficient of 0.1 is assumed for neutron energy 
depositions occurring in the plastic scintillator~\cite{nquench}.  Since the efficiency for rejecting coincident events with ZEPLIN--III
comes from detection of $\gamma$-rays, a large variation in quenching factor of the plastic scintillator makes negligible difference to the results presented.

Each energy deposition occurring within an active volume is converted to a number of
photoelectrons seen by the PMT. This has been performed in two ways. Firstly, an option in the Monte Carlo simulation allows
each of these energy depositions to be converted to scintillation photons, and for these scintillation photons
to be tracked to the photocathode, with losses occurring at points of 
reflection and due to the bulk attenuation. Rayleigh scattering is included. Poisson 
statistics are then applied to determine how many photoelectrons are generated. 
Since this is effectively a further complete simulation for each energy deposition, 
this option is slow and has been used only where necessary, principally to 
correlate measured TAL values with intrinsic BAL of the scintillators, and 
to allow the light emission of the scintillator to be determined.
The second much faster option is to simply apply the expected loss
due to the measured TAL given the distance from the end of the scintillator. 
Since the measured and simulated TALs agree, a further refinement to the simulation has been 
to use the Monte Carlo to estimate the fully 3D-dependent 
TAL throughout the scintillator, rather than a single mean value that 
describes the unit as a whole. 

\subsection{Overall Veto Performance}\label{ubersim}
\noindent The central function of the veto detector is to facilitate rejection of events that, on the basis of 
ZEPLIN--III data alone, could be mistaken for WIMPs. The principal characteristics of candidate
events in ZEPLIN--III are that they have an
electroluminescence-to-scintillation ratio characteristic of nuclear 
recoils, occur within the fiducial volume, and cause an energy deposition of $\leq$50~keVnr$^{1}$.
\footnotetext[1]{For a given event in the liquid xenon the (quenched) nuclear recoil energy 
may be determined from the scintillation signal, 
however, it is more convenient to calibrate the detector using electron recoils.  
This leads to the use of two energy scales: keVee (keV electron equivalent) and keVnr (keV nuclear recoil).  
The tradition in the field is to use 122~keV $\gamma$-rays from a $^{57}$Co source to set the energy calibration.}
The complete end-to-end simulation has been performed with 
all energy depositions in ZEPLIN--III and the veto recorded. While for the first science run 
the main contribution to background
was the activity of the PMTs of the ZEPLIN--III device itself, after the upgrade it is expected that no single source will dominate overwhelmingly - 
the contributions from PMTs, components and from the rock walls being similar. 
Differences arise due to the activity levels particular to each case, 
and due to the propagation of the radiation from point of origin through to detection.
Thus, the efficiency of the veto in rejecting background neutrons has been estimated by
considering each of these contributions 
separately, and then adding the contributions appropriately. 

\begin{figure}[ht]
%\vspace{5cm}
\includegraphics[width=8.6cm]{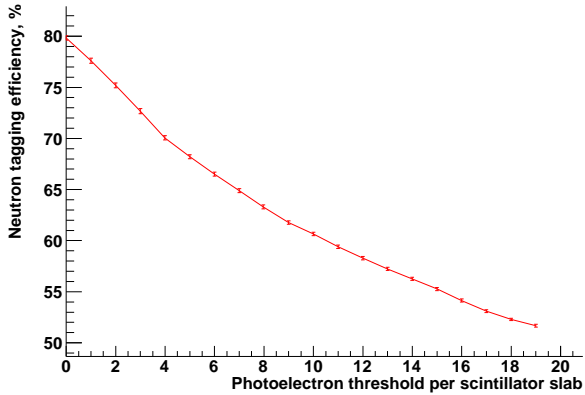}
\caption{\label{effvsphe}  Tagging efficiency of the veto for neutron events in ZEPLIN--III as a function of 
the signal size seen in at least one module of the veto, in terms of photoelectrons.}
\end{figure}

In its simplest mode, an event that satisfies the ZEPLIN--III criteria for a WIMP can be rejected by the veto 
if any one of the scintillator modules registers, within an appropriate time window ($-$20 to +300~$\mu$s), 
a signal significantly above background.  These are dubbed `tagged' events, 
and a tagging efficiency can be calculated as a function of the
veto signal size (see Figure~\ref{effvsphe}). 
A reasonably weak dependence on veto threshold is seen, with a maximum 
tagging efficiency at zero threshold of 80\% possible: about 20\% of neutron 
background events that satisfy the WIMP-search criteria in ZEPLIN--III 
have no interaction in any of the veto modules. 
In reality, a threshold of 6 photoelectrons in any single scintilator is sufficient 
to provide a minimal accidental coincidence contribution to overall event rate, and therefore this has been used as the 
conservative hard threshold within any module in 
simulations of the veto.  Lower limit efficiencies derived from this threshold determine the minimum neutron and gamma-ray background reduction in ZEPLIN--III.  
For the barrel modules, 6 photoelectrons corresponds to a typical energy deposition
of 135~keV at the far end and it is predicted that $>$65$\%$ of background neutrons that could otherwise have
been misidentified as WIMPs will be rejected by the veto with this threshold.  
However, as stated, since there is no hard threshold trigger for the veto 
when triggering in its slave mode, a higher tagging efficiency for coincident events (up to a maximum of 80\%) is achievable.    
This represents a substantial improvement on previous anti-coincidence systems such as the veto used by the ZEPLIN--II experiment~\cite{zep2si,zep2sd}, 
which had a neutron tagging efficiency of approximately 50\%, 
with a similar $\gamma$-ray tagging efficiency. 

\begin{figure}[ht]
\includegraphics[width=8.6cm]{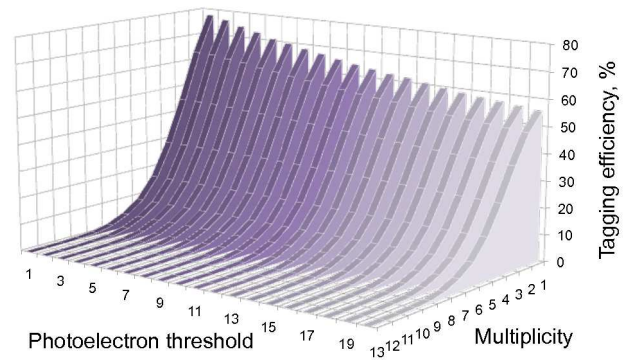}
\caption{\label{tagfrac} 
The efficiency for the veto to tag background neutron events that, based on ZEPLIN--III data alone, could
have been misidentified as WIMPs, as a function of the multiplivity of the event and the summed number of photoelectrons across all 
modules of the veto.  The multiplicity is defined as the number of modules 
contributing to the summed signal with at least one photoelectron.}
\end{figure}

Figure~\ref{effvsphe} shows the neutron background rejection power of the veto based on signals above
threshold in single scintillator modules only.  
In principle, it could be possible to 
increase the tagging efficiency further by searching for smaller energy depositions
occurring in multiple modules simultaneously.  Figure~\ref{tagfrac} extends  
Figure~\ref{effvsphe} to explore the summed energy deposition in multiple
scintillator modules. 
Since there exists no energy threshold for coincidence veto events when in slave mode of operation it is possible to 
apply not just single module thresholds, but a threshold on the sum energy of a veto event.  
In this case a veto event would be defined as the sum of all veto modules with signals above a minimal threshold as low as 
1 photoelectron and occuring in coincidence with one another within a short time window of a few $\mu$s duration.  The former condition 
takes advantage of the mutliple $\gamma$-rays emitted following Gd capture causing interactions in multiple modules to maximise the tagging 
efficiency of the veto system.  
The latter condition restricts the number of random coincidence events with ZEPLIN--III by demanding 
all $\gamma$-ray signals seen in multiple modules belong to the same event constrained within the timing window.  Since 320~$\mu$s timelines are 
recorded for each module 
a number of such events may be observed.  Figure~\ref{tagfrac} shows the neutron tagging efficiency of the veto as a function of the 
sum of photoelectrons detected 
from each modules registering at least one photoelectron.  It is seen that the tagging efficiency has a much stronger
dependence on the number of scintillator modules than on the signal size.   
With a summed signal of 6 photoelectrons the nominal 65\% tagging efficiency is acieved from any single module registering 6 or more photoelectrons.  
However, this efficiency is supplemented by events where the 6 photoelectrons are distributed, in any hit pattern, across multiple modules and the efficiency is increased to almost 70\%.  
The scope for improvement is limited to a maximum of approximately 80\% since, as seen in Figure~\ref{effvsphe}, some 20\% of candidate events result in no energy deposition 
in the veto at all.   
 
\begin{figure}[ht]
%\vspace{5cm}
\includegraphics[width=8.6cm]{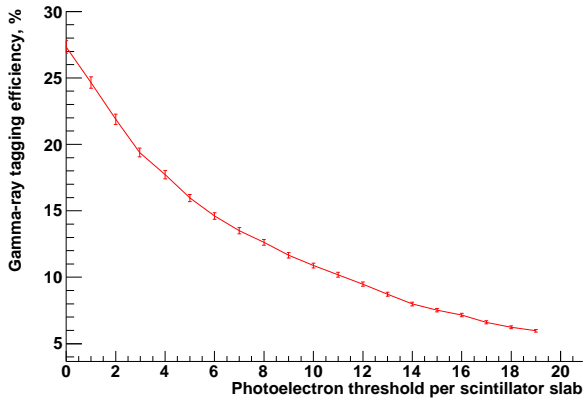}
\caption{\label{geffvsphe} $\gamma$-ray tagging efficiency of the veto as a function of the number of 
photoelectrons seen in at least one plastic scintillator.} 
\end{figure}

ZEPLIN--III has very good ability to discriminate between electron recoils, principally from $\gamma$-rays, and
nuclear recoils from neutrons or candidate WIMPs.  However, with sufficient exposure some $\gamma$-rays 
can still leak into the nuclear recoil acceptance region. 
Some fraction of these are likely to have also deposited energy in 
the veto, and thus could be rejected.  Simulations 
have been performed in which $\gamma$-rays were emitted from the known activities 
of the rock walls, laboratory, shielding, and from the detector components. 
Instances in which energy was deposited in both ZEPLIN--III (2--16~keVee, single scatter
in the fiducial volume) and in one or more of 
the veto plastic scintillators were searched for. The resulting efficiency for the veto to 
reject $\gamma$-rays that otherwise might have been misidentified as WIMPs, has been evaluated as a function
of the number of photoelectrons seen in at least one module of the 
veto, and is presented in Figure~\ref{geffvsphe}.
It is predicted that with a threshold of 6 photoelectrons being 
seen in a single module, $\sim$15\% of background
$\gamma$-rays depositing 2--16~keVee through single scatters in the ZEPLIN--III fiducial volume can be rejected by the veto.

\subsection{Optimisation of gadolinium loading}\label{gdloading}\label{pitchMC}

\noindent The adopted design relies heavily on the inclusion of gadolinium loaded into 
the polypropylene pieces of the veto.  The very high neutron capture cross section of 
$^{157}$Gd, combined with the subsequent emission of up to 8~MeV of $\gamma$-rays, is the
principle reason for the high neutron tagging efficiency.  Removing the Gd entirely results in the neutron tagging efficiency
for this geometry to drop to below 45\% for a 6 phe threshold.  
The results presented so far are for a loading of 0.5\% by weight, which was the value eventually adopted; here we explore the
effect of different concentrations and distributions of gadolinium. 

\begin{figure}[ht]
\includegraphics[width=8.6cm]{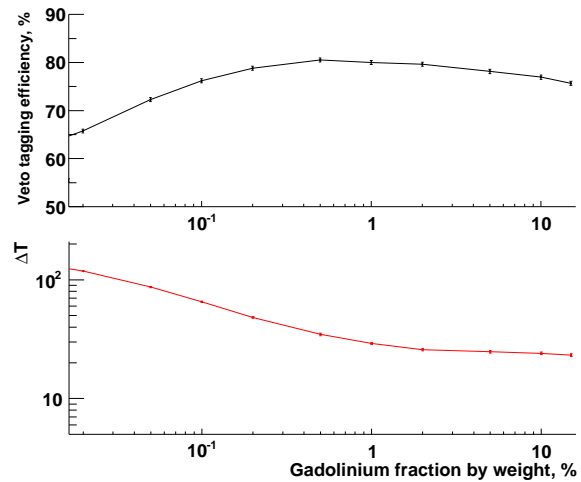}
\caption{\label{gdloadingfig} Monte Carlo simulations of the ZEPLIN--III -- veto system 
for various concentrations of gadolinium loading (by weight). Upper panel: the efficiency
for the veto to tag events that based on ZEPLIN--III data alone 
could have been misidentified as WIMPs.  Lower panel: For tagged events, the mean time delay in microseconds between
signals seen in ZEPLIN--III and the veto.}
\end{figure}

The incorporation of Gd into the polypropylene has two main effects. 
Firstly, increasing the fraction of gadolinium increases the fraction of internal neutrons that
are captured by the gadolinium isotopes, as opposed (mainly) to capture on the hydrogen
content of the polypropylene and plastic scintillator. 
Since the $^{157}$Gd capture in particular is followed by emission of several energetic $\gamma$-rays, 
this also tends to increase the tagging efficiency. This is illustrated in the upper panel of 
Figure~\ref{gdloadingfig} in which an exposure to neutrons from known background 
sources of the entire ZEPLIN--III -- veto system has been simulated.  For events that
satisfy the WIMP search criteria (fiducial volume, $\leq$50~keVnr) the fraction that 
result in 6 or more photoelectrons in a single module of the veto is shown, as a function
of the gadolinium loading. A gradual increase in tagging efficiency is seen up to about 
0.5\% fraction by weight. Above this little further gain is made, indicating that at this level of loading, essentially
all neutrons that thermalise in the polypropylene are captured by the gadolinium. 

The second effect of including gadolinium is that the mean time before neutron capture is reduced.
In terms of experimental infrastructure, a short capture time is desirable 
because it allows a larger fraction of coincidences to be
identified for a given coincidence time window, or alternatively, 
for a shorter window to be used
reducing data volume and the false-coincidence rate. 
The mean time between tagged signals in ZEPLIN--III and the veto, as a function of gadolinium concentration, 
is shown in the lower panel of Figure~\ref{gdloadingfig}. Here we see that
above about a 1\% loading, further increase ceases to reduce the 
capture time, indicating that capture is occurring very soon after thermalisation.  

Simulations have also shown that
neutrons are effectively moderated to thermal energies whether they originate externally to the veto (such as from the 
rock around the laboratory) or from within the internal volumes enclosed by the veto (such as from ZEPLIN--III itself).  External neutrons from the rock 
predominantly undergo radiative capture on H within the scintillator since very few are able to penetrate the 15~cm into the Gd-loaded polypropylene.  
The mean neutron energy at capture is 0.035~eV.
On the other hand, internal neutrons are captured on Gd and, at 0.5\% Gd concentration, the mean energy before capture is 0.045~eV (a slightly 
higher value due to the enhanced capture cross section relative to H).
In both cases, neutrons entering the veto are effectively
moderated and captured within the veto acquisition time window of 300~$\mu$s, with a mean time of 35~$\mu$s for the internal neutrons.  

\begin{figure}[ht]
%\vspace*{5cm}
\includegraphics[width=8.6cm]{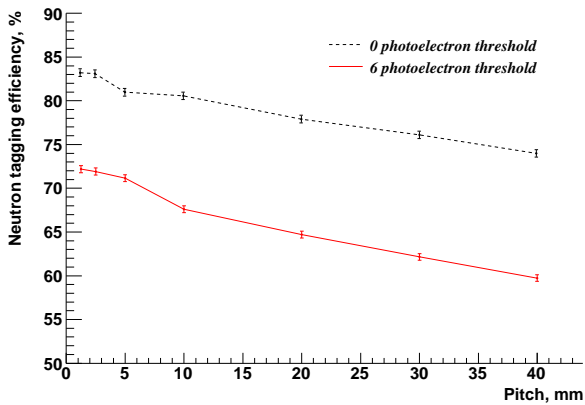}
\caption{\label{slots1} Monte Carlo simulation of the efficiency 
of the veto in rejecting events that, based on ZEPLIN--III data alone, 
could have been identified as WIMPs, as a function of the pitch of the Gd-loaded epoxy filled slots.    
The width of the slots, 2~mm, and the overall loading 
fraction of Gd, 0.5\% by weight, is maintained in all cases.}
\end{figure}

\begin{figure}[ht]
%\vspace*{5cm}
\includegraphics[width=8.6cm]{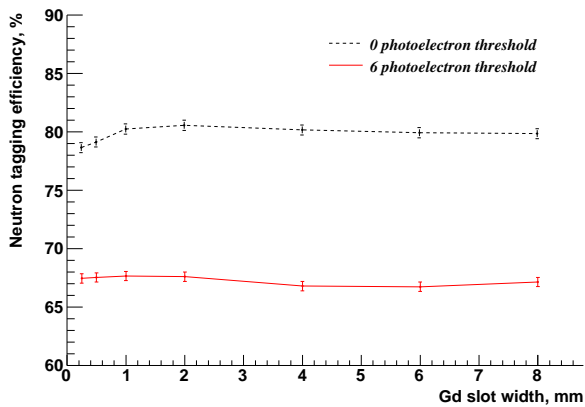}
\caption{\label{slots2}Monte Carlo simulation of the efficiency 
of the veto in rejecting events that, based on ZEPLIN--III data alone, 
could have been identified as WIMPs, as a function of the width of the Gd-loaded epoxy filled slots.  
The pitch of the slots, 10~mm, and the overall loading 
fraction of Gd, 0.5\% by weight, is maintained in all cases.}
\end{figure}

The gadolinium is suspended in epoxy that 
is located in 2~mm wide, 10~mm pitch slots that cut through the full depth of the polypropylene. This 
choice has partly been driven
by mechanical considerations, as narrower slots with a higher pitch would significantly increase the 
cost of machining. In Figures~\ref{slots1} and~\ref{slots2} the Monte 
Carlo simulation has been used to estimate 
the tagging efficiency as a function of slot width and slot pitch 
for a fixed average loading throughout the polypropylene of 0.5\%. Clearly, if the
slot pitch increases beyond about 10~mm, the efficiency of 
the veto in tagging coincident events within ZEPLIN--III 
begins to drop significantly. This is because, at this scale, 
neutrons moderating and thermalising within the polypropylene have 
a trajectory that sometimes never crosses a slot, and thus never provides an opportunity for
capture on the gadolinium loading. The width of the slots has less effect, 
and thus a value amenable to construction can be chosen that  
minimises cost and is optimal for the epoxy to hold the gadolinium in suspension.

\subsection{Predicted event rates}

\noindent %Here we present the total event rate that is expected in the veto detector. 
The large volume of plastic scintillator used in the veto, combined with a low threshold, has the potential to
create an extremely high event rate compatible neither with the data acquisition hardware nor
the coincidence methodology to be used.  
The thorough approach to control and evaluate the radioactivity content of items used, coupled to the
development of a detailed Monte Carlo simulation benchmarked against known performance, allows the absolute
event rates to be estimated. The dominant rate comes from $\gamma$-rays, although it is the neutron background generated by the veto that is of 
fundamental importance to ZEPLIN--III.  
Excluding any contributions from the veto, the net nuclear recoil background in the second science run of the ZEPLIN--III experiment is expected to be $\simeq$0.4 events per year.  
This is reduced to $\simeq$0.14 events per year for a 65\% veto tagging efficiency.  For each contribution of background from the veto the complete Monte Carlo was performed, with the neutrons and $\gamma$-rays distributed 
in energy and location as appropriate. Energy depositions in the veto and in ZEPLIN--III were recorded. 
The total expected event rate could then be calculated by summing the contributions.  
The net single scatter contribution to the background in the ZEPLIN--III acceptance region, 
assuming no rejection through vetoing the events, is 0.02 nuclear recoils per year.  
The nuclear recoil rate drops to 0.007 events per year when a conservative self-vetoing efficiency of 65\% is applied.  
The veto is predicted to contribute $<$1000 electron recoils per year from $\gamma$-ray emission within the ZEPLIN--III fiducial volume, in the energy range of 2--16 keVee.

\begin{figure}[ht]
%\vspace*{5cm}
\includegraphics[width=8.6cm]{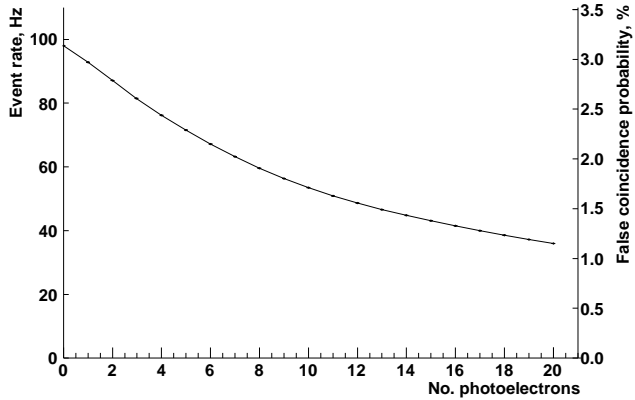}
%\vspace*{5cm}
\caption{\label{exposure}Predicted event rate of energy depositions in the veto,
as a function of signal size,  originating
from $\gamma$-ray emission of the measured activities of components.  Statistical errors, although present, are too small to be seen clearly on this scale.
Also shown is the probability for events seen in the veto following $\gamma$-ray emission from
components being misidentified as coincident events with ZEPLIN--III.}
\end{figure}

Considering the veto as a stand-alone instrument, Figure~\ref{exposure} shows the raw event rate that is expected in the veto as a
function of the number of photoelectrons observed in a single plastic scintillator module.
This shows, for example, that if a veto trigger requirement were imposed 
such that events were recorded if any single module recorded six or more photoelectrons, then a total 
event rate of 70~Hz would result.

We may consider how the background radioactivities will impact the veto when operating in 
coincidence with ZEPLIN--III by calculating the `accidental coincidences rate'.  This is the probability of 
any recorded timeline having an uncorrelated background-radioactivity caused signal within it.  
For the usual mode of slave operation, with the veto triggered to record for 320~$\mu$s by events
from ZEPLIN--III at an estimated rate of~1~Hz, 
the accidental coincidences rate is indicated by the right hand axis in Figure~\ref{exposure}.
For nominal operational parameters the rate is less than 2\%.

\subsection{Impact of gaps in shielding}\label{voidsMC}

\noindent The veto instrument consists of many modular pieces mechanically held together. As described in 
section~\ref{overalldesign}, the plastic scintillators are deliberately offset relative to
the polypropylene modules to prevent line-of-sight gaps that would compromise the shielding. There is, 
however, still a significant chance of introducing small gaps during construction and assembly. 
The Monte Carlo simulation has been employed to estimate the impact of 
such gaps occurring between the active modules, both in terms of 
degrading the shielding from external sources of background, and in terms of lowering
the ability of the veto to tag coincident nuclear recoil events. 

The simulation was performed for two distinct cases: when gaps existed 
between both active and passive modules, and when gaps existed 
between only active modules (in which case the passive modules were assumed to be
completely closed). 
In the former case a direct path is formed for radiation that has penetrated the
external lead shield to reach ZEPLIN--III. %Consequently, the neutron flux incident on
%ZEPLIN--III is increased.  However, 
Simulation of a three year exposure that included gaps 
of 1~mm width, in excess of engineering constraints, showed no statistically significant 
increase in the ZEPLIN--III active volume single scatter rate. This is consistent with 
previous simulations performed to explore the impact of shielding due to the 
existing holes needed for pipework and cabling.
Moreover, the veto tagging efficiency remains unaffected until the gaps exceed a width of 1~cm.  
This is because a neutron, having originally entered the shielding or been internally generated, is 
unlikely to interact with ZEPLIN--III and exit the shielding through a gap again.  
In the latter sets of simulations, in which gaps between adjacent active
scintillator modules were introduced assuming a continuous inner Gd loaded polypropylene barrel, it was found that 
the efficiency for tagging coincident veto -- ZEPLIN--III nuclear recoil events 
begins to be significantly reduced for gaps of widths greater than 2~cm, essentially due to less 
coverage for $\gamma$-ray detection following neutron capture.  The neutron flux 
seen by ZEPLIN--III from external background is 
unaffected until the gap exceeds 1~cm between scintillator modules.

\section{Present Status}
\noindent 
All veto components have been shipped underground to the Boulby Laboratory with active and passive modules coupled in preparation for installation.  
The full veto array, complete with data acquisition and synchronisation hardware, LED calibration systems and data reduction software 
has been successfully assembled within the Pb castle.  This was done in the absence of ZEPLIN--III to test systems prior to final installation.  Neutron and $\gamma$-ray source exposures as well as background runs have
been performed and data recorded for {\em in situ} calibration and characterisation of the detector.  The veto has since been dismantled to allow access to ZEPLIN--III. 
It will be installed following the completion of the upgrade to the ZEPLIN--III photomultiplier array.  
Performance results of the veto will be presented following commencement of the second science run of the ZEPLIN--III experiment.  

\section{Summary}

\noindent 
The second science run of the ZEPLIN--III project will feature two main upgrades: lower background 
photomultiplier tubes and the use of an active veto. Here, the design of the veto has been presented,
together with details of the radiological content and performance of the components to be used.
Detailed GEANT4 Monte Carlo simulations have been used to aid in the characterisation of
the veto, and to estimate its overall performance. In addition to providing valuable diagnostic information, it is expected that the veto will be able to 
reject over 65\% of neutrons, and over 15\% of $\gamma$-rays from background radioactivities, whilst
contributing negligibly to the ZEPLIN--III acceptance region background, expected to be $\simeq$0.4 events per year.  The veto will reduce this background to less than 
$\simeq$0.14\ events per year.  This is a significant factor in the event of a non-zero observation.
The veto has been fully assembled and systems integrated, and will be installed around ZEPLIN--III shortly for the 
commencement of the second science run as this instrument probes yet greater sensitivity.

\begin{acknowledgments}

The UK groups acknowledge the support of the Science \& Technology Facilities Council (STFC) for the ZEPLIN--III project and for maintenance and operation of the underground Palmer 
laboratory which is hosted by Cleveland Potash Ltd (CPL) at Boulby Mine, near Whitby on the North-East coast of England.  
The project would not be possible without the co-operation of the management and staff of CPL. 
We also acknowledge support from a Joint International Project award, held at ITEP and ICL, from the Russian Foundation of Basic Research (08-02-91851 KO a) and the Royal Society.
LIP--Coimbra acknowledges financial support from Funda\c c\~ao para a Ci\^encia e Tecnologia (FCT) through the project-grants CERN/FP/83501/2008 and CERN/FP/109320/2009, 
as well as the  postdoctoral grants SFRH/BPD/27054/2006 and SFRH/BPD/47320/2008.  
This work was supported in part by SC Rosatom; by Russian Grant  SS-1329.2008.2 and by the Russian Ministry of Education and Science contract 02.740.11.0239.  
The University of Edinburgh is a charitable body, registered in Scotland, with the registration number SC005336.

\end{acknowledgments}

\end{document}